\documentclass[11pt]{article}

\usepackage{arxiv}
\usepackage[T1]{fontenc}
\usepackage[utf8]{inputenc}
\IfFileExists{newtxtext.sty}%
  {\usepackage{newtxtext}\usepackage[varvw]{newtxmath}}%
  {\usepackage{mathptmx}\usepackage[scaled=0.9]{helvet}\usepackage{amssymb}}
\usepackage{textcomp}
\usepackage{microtype}
\usepackage{amsmath}
\usepackage{graphicx}
\graphicspath{{figures/}}
\usepackage{cite}
\usepackage[labelfont=bf,labelsep=space,justification=justified,singlelinecheck=false,font=small]{caption}
\newcommand{\captiontitle}[1]{\textbf{#1}}
\usepackage{xcolor}
\usepackage[colorlinks=true,linkcolor=blue!50!black,citecolor=blue!50!black,urlcolor=blue!50!black]{hyperref}
\hypersetup{pdftitle={BIC-Engineered Silicon Nitride Microring Resonators Enabled by an Etch-Free Laser-Writing Approach},
            pdfauthor={L. Hajshahvaladi, M. A. Salmani, M. Abbasi, M. Yaghoobi, M. Asgari, G.-M. Parsanasab}}

\title{BIC-Engineered Silicon Nitride Microring Resonators\\Enabled by an Etch-Free Laser-Writing Approach}
\shorttitle{BIC-Engineered Silicon Nitride Microring Resonators}
\headauthor{Hajshahvaladi \textit{et al.}}
\date{November 2026}

\author{%
  Leila Hajshahvaladi, \ Mohammad Amin Salmani, \ Media Abbasi, \ Mobin Yaghoobi,\\
  Mohammad Asgari, \ Gholam-Mohammad Parsanasab\textsuperscript{*}\\[4pt]
  \small Integrated Photonics Laboratory (IPL), Faculty of Electrical Engineering\\
  \small Shahid Beheshti University, Tehran, Iran\\[3pt]
  \small \textsuperscript{*}Corresponding author: \href{mailto:gm_parsanasab@sbu.ac.ir}{gm\_parsanasab@sbu.ac.ir}}

\begin{document}
\maketitle

\begin{abstract}
Silicon nitride photonics rests on well-established fabrication processes that deliver waveguides and resonators of exceptionally low loss. Here we report an initial experimental demonstration of a complementary, etch-free approach with fewer fabrication steps, in which SU-8 ridges written by direct ultraviolet laser lithography on an unpatterned silicon nitride film define waveguides and microring resonators. Because the nitride film is never patterned, scattering at patterned nitride sidewalls does not arise, and the remaining radiative loss is engineered by design. The guided transverse-magnetic mode lies within the continuum of transverse-electric slab modes, and we use bound states in the continuum (BICs) as a design tool to control this radiation. Coupled-mode analysis and full-vectorial simulations show that radiation from the two ridge sidewalls cancels by destructive interference at discrete widths, with a first BIC at 2.25~\textmu m and radiation below 1~dB/cm over a 0.3~\textmu m width window and about 200~nm of bandwidth, and that curvature confines this cancellation to discrete bands of bending radius and width. Laser-written waveguides follow the predicted width dependence of the loss, reaching approximately 1.5~dB/cm near the BIC, where non-radiative loss dominates. A microring designed on a radius-selected BIC band exhibits loaded quality factors up to 1.6 × 10\textsuperscript{4} and, under the over-coupled assignment supported by a coupler model, an intrinsic quality factor of 3.3 × 10\textsuperscript{5}. Because radiation is set by a patterned low-index layer on an unetched film, the same approach could be extended to other nanostructures. Optimization of the writing conditions and materials is ongoing.
\end{abstract}

\begin{keywords}
silicon nitride photonics \textperiodcentered{} bound states in the continuum \textperiodcentered{} direct laser lithography \textperiodcentered{} microring resonators \textperiodcentered{} etch-free waveguides \textperiodcentered{} destructive interference
\end{keywords}

\section{Introduction}\label{sec:introduction}

Silicon nitride (Si\textsubscript{3}N\textsubscript{4}) has become a leading platform for integrated photonics~\cite{bogaerts,pelucchi,elshaari} owing to its broad transparency window, weak nonlinear absorption and compatibility with low-loss, dispersion-engineered waveguides~\cite{liu2021,xiang2021,churaev}. These properties have enabled frequency combs~\cite{chang2022,pfeiffer2017}, narrow-linewidth lasers~\cite{xiang2021,corato}, nonlinear frequency conversion~\cite{lu2021,nitiss}, microwave photonics~\cite{marpaung} and quantum light sources~\cite{li2025,mahmudlu}. These achievements rest on well-established fabrication processes, in which the nitride is patterned by lithography and dry etching. In patterned waveguides, the residual loss is set largely by scattering at the patterned boundaries~\cite{sorayaie,pfeiffer2018}, and dedicated techniques such as Damascene reflow and high-temperature annealing have reduced it to the point where intrinsic quality factors exceed \(10^{8}\)~\cite{pfeiffer2018,pfeiffer2016,puckett}. Alongside these processes, approaches that require fewer fabrication steps are of interest for rapid prototyping and for laboratories that do not operate wafer-scale nanofabrication facilities.

One such approach is to leave the high-index film unpatterned and define the waveguide in a fabrication-friendly, low-index material on top of it. The patterned ridge raises the local effective index and confines light laterally in the underlying film, so the functional layer is never etched~\cite{zou2015,yu2019}. This hybrid architecture has been established mainly on thin-film lithium niobate with electron-beam-patterned resists~\cite{yu2019,yu2020nc,yu2020lsa,feng2023}. It has since been extended to two-dimensional materials~\cite{yu2019adom}, near-visible wavelengths~\cite{yu2021adom}, rare-earth oxides~\cite{xu2021}, semi-nonlinear Si\textsubscript{3}N\textsubscript{4}/LiNbO\textsubscript{3} waveguides~\cite{li2022} and heterogeneous Si\textsubscript{3}N\textsubscript{4}/LiNbO\textsubscript{3} circuits~\cite{han2024}. Its price is radiation. The laterally confined mode generally has a lower effective index than a slab mode of orthogonal polarization, so it lies within the slab continuum and leaks into it through polarization conversion at the ridge sidewalls. Low loss therefore requires a geometry for which this leakage cancels, a condition known as a bound state in the continuum (BIC)~\cite{hsu2016,hsu2013,doeleman}. Photonic BICs can arise from symmetry protection, from Friedrich--Wintgen interference between resonances~\cite{fw1985} or from parameter-tuned (accidental) cancellation between radiation pathways~\cite{hsu2016,bulgakov}. The last mechanism makes radiation a continuous function of structural parameters that can be set by design~\cite{kang2023}. Waveguide BICs of this type have been studied extensively~\cite{zou2015,yu2019,feng2023,han2024,gu2023,ctyroky2023}, and BIC concepts are being applied to Si\textsubscript{3}N\textsubscript{4} microresonators~\cite{lima2026}. Two aspects remain open for an accessible Si\textsubscript{3}N\textsubscript{4} platform based on this architecture. The first is fabrication. Previous hybrid BIC devices were patterned by electron-beam lithography, which provides high resolution. Maskless direct ultraviolet laser lithography of a photo-patternable polymer such as SU-8 requires less equipment and fewer steps, but its coarser resolution and larger dimensional variation must be accommodated by the optical design. Our laboratory has previously fabricated SU-8 polymer microring lasers and resonators by direct laser lithography~\cite{sorayaie,hajshahvaladi2025,kavousi2025}, and the present work extends this approach to Si\textsubscript{3}N\textsubscript{4}. The second aspect is the resonator. Curvature shifts the mode towards the outer sidewall and unbalances the two radiation pathways, so the straight-waveguide BIC no longer guarantees low loss, and the ring--bus coupler adds an external loss channel and a second radiating structure. Previous BIC ring and spiral designs used the straight-waveguide condition with semi-analytical bend corrections~\cite{yu2019} or radii large enough to make bending negligible~\cite{yuan2023}. A compact resonator instead requires the width, radius and coupler to be designed together.

Here we report hybrid SU-8/Si\textsubscript{3}N\textsubscript{4} waveguides and microring resonators fabricated by direct UV laser lithography on an unetched nitride film, together with the numerical design that makes them low-loss. The SU-8, applied by spin coating, serves both as the photoresist and as the waveguide material, so a single maskless exposure and development define the complete waveguide geometry without additional thin-film deposition. Because the nitride film is never patterned, scattering at patterned nitride sidewalls does not arise, and the loss channels that remain are radiation into the film, which the design controls, and non-radiative loss in the materials and at the polymer ridge. We do not present the BIC phenomenon itself as the novelty of this work. Rather, we use it as a design tool that makes this etch-free route usable, by selecting the ridge width, bending radius and coupler geometry, and we use the measured loss and resonator response to verify that the fabricated structures reproduce the intended modal behavior. The approach has clear limitations: guidance is restricted to the TM-like mode, the weak lateral confinement requires bend radii of the order of 100~\textmu m, and the present losses are higher than those achieved with mature Si\textsubscript{3}N\textsubscript{4} processes. We therefore present it as an additional route with fewer fabrication steps, complementary to established processes rather than a replacement for them. Figure~\ref{fig:1} introduces the concept and the interference mechanism, Fig.~\ref{fig:2} characterizes the width selection of the BIC and its behavior in a real device, Fig.~\ref{fig:3} extends the design to bends and couplers, and Fig.~\ref{fig:4} presents the fabricated microring.

\section{Results}\label{sec:results}

\subsection{Hybrid platform and its radiation channel}\label{sec:hybrid-platform-and-its-radiation-channe}

The platform consists of SU-8 ridge waveguides and microrings patterned on an unetched Si\textsubscript{3}N\textsubscript{4} film; in the resonator geometry, a ring of radius \(R\) is side-coupled through a gap \(g\) to a bus waveguide [Fig.~\ref{fig:1}(a)]. SU-8 plays a dual role. As a negative photoresist it is patterned directly by the laser, and as a transparent, low-index dielectric it remains as the permanent ridge that confines light in the nitride film beneath. The ridge height, set by spin coating, and the ridge width, set by the exposure dose, are therefore the two geometric parameters defined in fabrication. The cross-section comprises an 800~nm-thick SU-8 ridge of width \(w\) on a 400~nm-thick Si\textsubscript{3}N\textsubscript{4} layer, a 2~\textmu m SiO\textsubscript{2} buffer on silicon and an air upper cladding [Fig.~\ref{fig:1}(b)]; the model uses refractive indices of 1.58 (SU-8), 2.15 (Si\textsubscript{3}N\textsubscript{4}) and 1.45 (SiO\textsubscript{2}) at \(\lambda = 1550\)~nm. The two sidewalls of the ridge act as coherent radiation channels, and the interference between them [Fig.~\ref{fig:1}(c)], analyzed in Section 2.2, is the central element of the design.

The ridge laterally confines a TM-like guided mode, whose dominant electric field \(E_{y}\) is normal to the film and concentrated in the Si\textsubscript{3}N\textsubscript{4} beneath the ridge, with effective index \(n_{\mathrm{TM}} = 1.742\) at \(w = 2.25\)~\textmu m [Fig.~\ref{fig:1}(b)]. Outside the ridge, the film supports a TE slab mode, with in-plane electric field, of higher effective index \(n_{\mathrm{TE}} = 1.858\), because the in-plane field of the TE mode is continuous across the film surfaces whereas the normal field of the TM mode is not. A TE slab wave can travel in the film plane at any angle \(\theta\) to the waveguide axis, with axial propagation constant \(k_{0}n_{\mathrm{TE}}\cos\theta\), where \(k_{0} = 2\pi/\lambda\), so the TE waves form a continuum that covers every axial effective index below \(n_{\mathrm{TE}}\) [Fig.~\ref{fig:1}(d)]. Since \(n_{\mathrm{TM}} < n_{\mathrm{TE}}\), the guided mode is spectrally embedded in this continuum: at its propagation constant \(\beta = k_{0}n_{\mathrm{TM}}\) it is phase-matched to the TE waves of lateral wavenumber \(\pm k_{x,c}\), \(k_{x,c} = k_{0}\left( n_{\mathrm{TE}}^{2} - n_{\mathrm{TM}}^{2} \right)^{1/2} \approx 2.62\)~\textmu m$^{-1}$, which travel at \(\theta = \arccos\left( n_{\mathrm{TM}}/n_{\mathrm{TE}} \right) \approx 20^{\circ}\) to the axis. The TM slab mode outside the ridge has an effective index below \(n_{\mathrm{TM}}\), which is why the mode is laterally guided, so the TE slab continuum is the only open radiation channel. Phase matching to this continuum is available at every width, but it is a necessary rather than a sufficient condition for radiation. Whether the guided mode radiates is set by its coupling to the phase-matched TE waves, and because the mode is protected neither by total internal reflection nor by a band gap, low loss requires this coupling to vanish. Figures~\ref{fig:1}(e) and 1(f) show the simulated fields of the TM-like guided mode and of a TE slab continuum mode at \(w = 2.25\)~\textmu m.

\subsection{Two-sidewall interference and the BIC condition}\label{sec:two-sidewall-interference-and-the-bic-co}

\begin{figure}[!tbp]
  \centering
  \includegraphics[width=\textwidth]{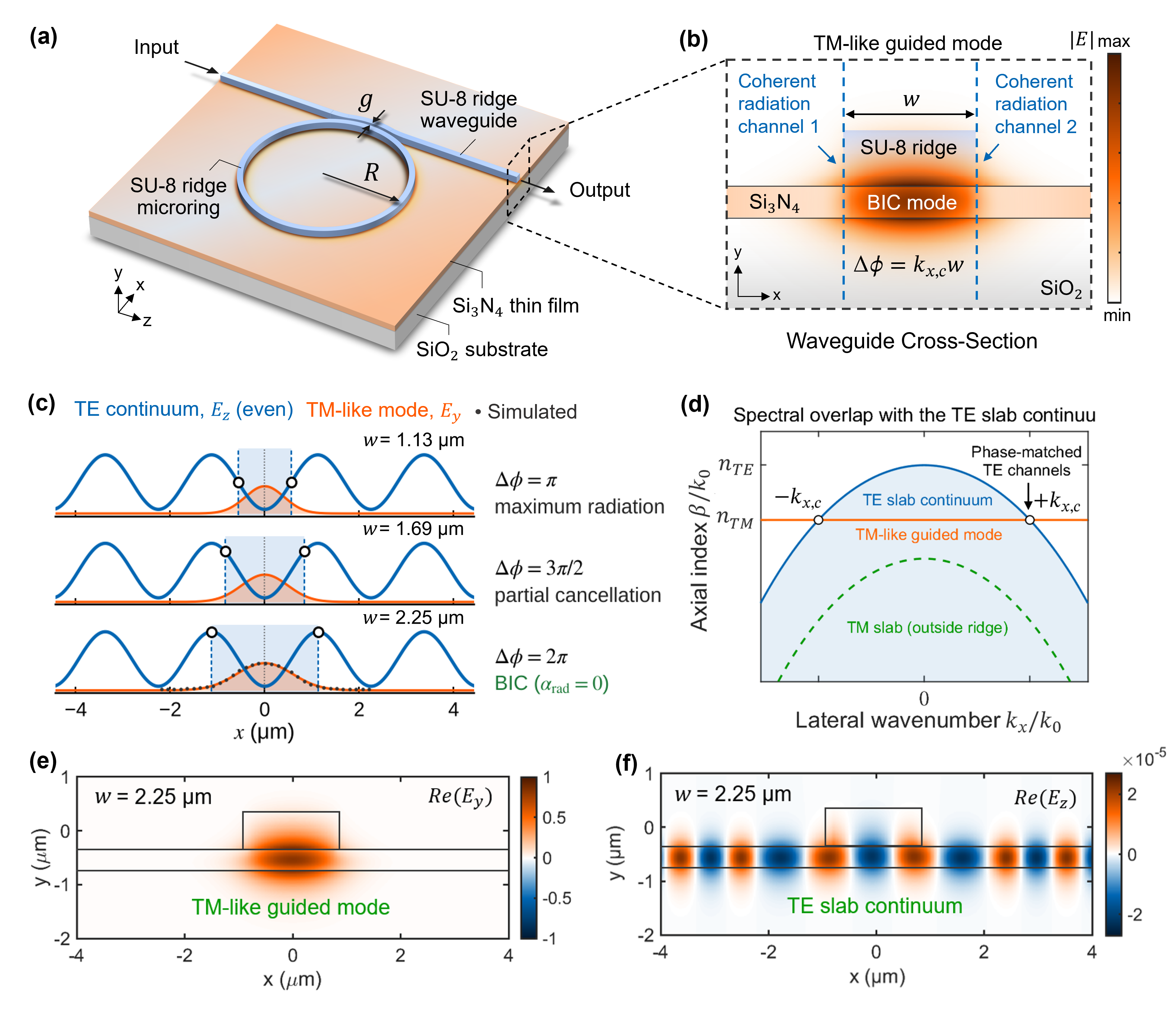}
  \caption{\captiontitle{Concept: a BIC-guided waveguide on an unetched Si\textsubscript{3}N\textsubscript{4} film.} \textbf{(a)} SU-8 ridge microring of radius \(R\), side-coupled through a gap \(g\) to an SU-8 bus waveguide, written by direct laser lithography on an unetched Si\textsubscript{3}N\textsubscript{4} film on SiO\textsubscript{2}/Si. \textbf{(b)} Waveguide cross-section with the simulated field magnitude \(|E|\) of the TM-like guided mode at the BIC width (labeled BIC mode). The SU-8 sidewalls at \(x = \pm w/2\) (dashed) are the only locations where the guided mode converts into TE slab waves; they act as two coherent radiation channels whose waves differ in phase by \(\Delta\phi = k_{x,c}w\). \textbf{(c)} Two-sidewall interference for three ridge widths, on the lateral scale of (e) and (f). Orange: \(E_{y}\) of the TM-like guided mode (points: simulated profile of (e)). Blue: \(E_{z}\) of the even TE continuum mode. Circles mark the sidewalls. The waves launched at the two sidewalls add at \(\Delta\phi = \pi\) (maximum radiation), partly cancel at \(3\pi/2\) and cancel completely at \(2\pi\), where \(\alpha_{\mathrm{rad}} = 0\): the BIC. \textbf{(d)} Spectral overlap, shown as an in-plane wavevector diagram at fixed frequency. The TE slab continuum (blue) covers all axial indices below \(n_{\mathrm{TE}}\), so the TM-like guided mode (orange line) is phase-matched to it at the lateral wavenumbers \(\pm k_{x,c}\); the TM slab mode outside the ridge (dashed) lies below \(n_{\mathrm{TM}}\). \textbf{(e)}, \textbf{(f)} Simulated fields at \(w = 2.25\)~\textmu m: \(\mathrm{Re}\left( E_{y} \right)\) of the TM-like guided mode (e) and \(\mathrm{Re}\left( E_{z} \right)\) of a TE slab continuum mode (f). Both are symmetric under \(x \rightarrow - x\), that is, both modes are even. Panels (b), (e) and (f) are full-vectorial simulations; (c) and (d) are model illustrations; (a) is a schematic. Parameters: SU-8 thickness 800~nm, Si\textsubscript{3}N\textsubscript{4} thickness 400~nm, \(\lambda = 1550\)~nm; refractive indices 1.58 (SU-8), 2.15 (Si\textsubscript{3}N\textsubscript{4}) and 1.45 (SiO\textsubscript{2}).}
  \label{fig:1}
\end{figure}

We describe the ridge as a perturbation \(\Delta\varepsilon(x,y)\) of the unloaded layer stack, equal to \(n_{\mathrm{SU8}}^{2} - 1\) within the ridge cross-section and zero elsewhere, and expand the field in the eigenmodes of the stack: the TM-like guided mode with amplitude \(a(z)\) and the TE slab waves with amplitudes \(c\left( k_{x},z \right)\), normalized to unit power per unit lateral wavenumber~\cite{zou2015,han2024}. Two properties of the structure determine the coupling between them. First, within any laterally uniform region the local TM and TE eigenmodes are orthogonal, so polarization conversion requires a lateral variation of the permittivity and is confined to the SU-8 sidewalls, where \(\partial\Delta\varepsilon/\partial x\) is non-zero. Second, conversion at a sidewall requires vertical asymmetry. In a stack that is symmetric under \(y \rightarrow - y\), the TE and TM slab modes have opposite parity and remain uncoupled at any lateral step; here the SU-8 ridge above the film and the SiO\textsubscript{2} buffer below it break this symmetry, so each sidewall acts as a source of TE radiation. The coupling to a TE wave of lateral wavenumber \(k_{x}\) is therefore the lateral Fourier component of a source localized at the sidewalls, \(\widetilde{\kappa}\left( k_{x} \right) = \int\sigma(x)\, e^{- ik_{x}x}\, dx\) with \(\sigma(x) = \sigma_{0}\left\lbrack \delta(x + w/2) - \delta(x - w/2) \right\rbrack\). The opposite signs follow from the opposite orientations of the index step, \(\partial\Delta\varepsilon/\partial x\), at the two sidewalls, whereas the laterally even guided mode has the same amplitude at both. The single-sidewall strength \(\sigma_{0}\) (units of m$^{-1/2}$) combines the index step \(n_{\mathrm{SU8}}^{2} - 1\), the vertical overlap of the TM-like and TE profiles across the ridge height, and the guided-mode amplitude at the sidewall. For the phase-matched channel, \(\widetilde{\kappa}\left( \pm k_{x,c} \right) = \pm 2i\sigma_{0}\sin\left( k_{x,c}w/2 \right)\): the waves launched at the two sidewalls accumulate a relative phase \(\Delta\phi = k_{x,c}w\) across the ridge, and their superposition is governed by this phase [Fig.~\ref{fig:1}(b), (c)].

The two modes involved share the same mirror symmetry. The cross-section is symmetric about \(x = 0\), and the TM-like guided mode is even under \(x \rightarrow - x\). The phase-matched TE waves \(\pm k_{x,c}\) combine into standing waves of definite symmetry. The even TE slab mode has a symmetric longitudinal field, \(E_{z} \propto \cos\left( k_{x,c}x \right)\), as in the simulated continuum mode [Fig.~\ref{fig:1}(f)], and an antisymmetric transverse field \(E_{x} \propto \sin\left( k_{x,c}x \right)\). This mode is the radiation channel of the waveguide: it has the symmetry of the guided mode, so their coupling is allowed at every width, and its magnitude, \(2\left| \sigma_{0}\sin\left( k_{x,c}w/2 \right) \right|\), is set by the interference of the two sidewall contributions [Fig.~\ref{fig:1}(c)]. The odd TE standing wave has the opposite symmetry and takes no part in the radiation. Both behaviors appear in the eigenmodes of a laterally finite slab, in which the continuum becomes a discrete ladder of TE modes whose effective indices sweep through \(n_{\mathrm{TM}}\) as the slab width varies: even TE branches anti-cross the guided mode, showing that the even channel is open, whereas odd TE branches cross it without splitting [Fig.~\ref{fig:2}(a)].

The BIC therefore originates from destructive interference, not from symmetry protection. In a symmetry-protected BIC, the bound mode and the continuum have opposite symmetry and are decoupled at every width. Here the guided mode and its radiation channel are both even, and the guided mode generally radiates [Fig.~\ref{fig:1}(c), upper and middle rows]. Radiation vanishes only where the waves launched at the two sidewalls cancel, at \(\Delta\phi = k_{x,c}w = 2\pi m\). The sidewalls then sit at extrema of \(E_{z}\) of the even TE mode, where its transverse component \(E_{x}\), normal to the sidewalls, vanishes [Fig.~\ref{fig:1}(c), lower row]. The BIC is thus an interference (accidental) BIC, reached by tuning the ridge width, and it disappears when the width departs from this condition.

Eliminating the continuum amplitudes in the Wigner--Weisskopf (Markov) approximation, which holds when \(\widetilde{\kappa}\) varies slowly over the range of \(k_{x}\) phase-matched within a propagation length, gives a Fermi-golden-rule decay of the guided power, \(\left| a(z) \right|^{2} \propto e^{- \alpha_{\mathrm{rad}}z}\)~\cite{zou2015,hsu2016}. Each channel contributes \(2\pi\left| \widetilde{\kappa}\left( k_{x,c} \right) \right|^{2}\,\left| dk_{x}/d\beta \right|\). For the one-dimensional lateral continuum of the film, \(\beta^{2} + k_{x}^{2} = k_{0}^{2}n_{\mathrm{TE}}^{2}\) gives \(\left| dk_{x}/d\beta \right| = \beta/k_{x,c} = \cot\theta\), which expresses the growing density of TE states as the phase-matched wave approaches grazing propagation. Summing the two channels, which couple with equal strength, yields the radiation loss of the guided mode into the TE slab continuum:

\begin{equation}
  \alpha_{\mathrm{rad}}(w) = \frac{16\pi\, n_{\mathrm{TM}}}{\sqrt{n_{\mathrm{TE}}^{2} - n_{\mathrm{TM}}^{2}}}\,\left| \sigma_{0}(w) \right|^{2}\,\sin^{2}\left( \frac{k_{x,c}w}{2} \right).
  \label{eq:loss}
\end{equation}
Here \(\alpha_{\mathrm{rad}}\) is the power attenuation coefficient; expressed in m$^{-1}$, it corresponds to a loss of \(0.0434\,\alpha_{\mathrm{rad}}\) in dB/cm. The prefactor \(16\pi n_{\mathrm{TM}}/\left( n_{\mathrm{TE}}^{2} - n_{\mathrm{TM}}^{2} \right)^{1/2} = 16\pi\cot\theta\) is dimensionless, so \(\alpha_{\mathrm{rad}}\) carries the units of \(\left| \sigma_{0} \right|^{2}\). The width enters through the interference factor and, more weakly, through \(\sigma_{0}(w)\), whose guided-mode amplitude at the sidewall falls as the lateral mode broadens. Equation~(\ref{eq:loss}) assumes that the ridge is a weak perturbation, so the coupling can be evaluated to first order; that the sidewalls are abrupt vertical steps; that leakage through the 2~\textmu m buffer into the silicon substrate is negligible; and that absorption and scattering are excluded, so \(\alpha_{\mathrm{rad}}\) is the purely radiative loss computed in the full-vectorial simulations.

Because the prefactor of Eq.~(\ref{eq:loss}) is positive, radiation vanishes only when the interference factor does, at \(k_{x,c}w_{m} = 2\pi m\) (\(m = 1,2,\ldots\)), where the wave launched at one sidewall reaches the other with a phase delay of \(2\pi m\) and cancels the oppositely signed wave launched there. Only one channel, the even TE slab mode, has to be cancelled, and because time-reversal symmetry allows its coupling constant to be chosen real, the cancellation is a single real condition~\cite{hsu2016}. The BIC is therefore accidental yet structurally stable: a small change of parameters displaces it along a one-parameter family, a line in the width--wavelength plane, rather than removing it.

\subsection{Width selection and experimental validation in straight waveguides}\label{sec:width-selection-and-experimental-validat}

Figure~\ref{fig:2}(b) tests this picture quantitatively. Both the analytical model [Eq.~(\ref{eq:loss})] and full-vectorial eigenmode simulations show radiation losses that fall by more than three orders of magnitude, from 10--100~dB/cm to below 0.01~dB/cm, at \(w \approx 2.25\)~\textmu m and again at \(w \approx 4.6\)~\textmu m, separated by maxima of about 15~dB/cm near 3.3~\textmu m and 4~dB/cm near 5.8~\textmu m. Orthogonality of polarization or symmetry would suppress radiation at all widths. Instead, the loss vanishes only at periodic widths whose spacing, about 2.35~\textmu m, agrees with the lateral period \(2\pi/k_{x,c} \approx 2.4\)~\textmu m, which identifies two-sidewall interference as the mechanism. The decrease of the maxima reflects the envelope of Eq.~(\ref{eq:loss}), that is, the weaker field of a laterally broader mode at the sidewalls. Evaluated with the effective-index lateral mode \(f_{b}(x)\) (Section 2.4 and Methods), the envelope \(\left| f_{b}(w/2) \right|^{2}\cot\theta\) predicts a ratio of 4.0 between the maxima at 3.3 and 5.8~\textmu m, close to the full-vectorial ratio of 3.75 and steeper than a pure \(w^{- 2}\) law (3.1) [Fig.~\ref{fig:2}(c)]. The model fixes the period and functional form of the loss but not the absolute BIC position: with the full-vectorial indices, \(k_{x,c}w = 2\pi\) gives \(w = 2.4\)~\textmu m, about 6\% above the simulated value, which we attribute to the replacement of sidewalls of finite height by abrupt lateral steps.

The triangles in Fig.~\ref{fig:2}(b) show propagation losses of straight laser-written waveguides measured by the top-scattering method (Methods). Their width dependence reproduces the predicted interference pattern. Between \(w \approx 2.6\) and 4.3~\textmu m, the measured losses of 4--15~dB/cm follow the simulated radiation loss, showing that radiation into the TE slab continuum dominates away from the BIC. The lowest measured loss, approximately 1.6~dB/cm, occurs at \(w \approx 2.3\)~\textmu m, coincident with the first predicted BIC. Here the measurement exceeds the radiation-only prediction by about two orders of magnitude: once radiation is suppressed, absorption in SU-8 and Si\textsubscript{3}N\textsubscript{4} and scattering at the laser-written sidewalls and the SU-8/Si\textsubscript{3}N\textsubscript{4} interface, none of which are included in the simulation, set the loss floor (see Discussion). The minimum thus marks the crossover from a radiation-limited to a non-radiatively limited regime, the experimental signature expected of a BIC in a waveguide with finite material and fabrication losses~\cite{yu2019,han2024}. For \(w \lesssim 2\)~\textmu m, several measured values lie below the simulated radiation loss. Because the radiation-only simulation is a lower bound for the nominal cross-section, these points cannot be explained by additional loss mechanisms. Two factors contribute: the width assigned to each waveguide, which is critical in this range because the predicted loss changes by more than an order of magnitude over 0.3~\textmu m, and a possible underestimate of large losses by the scattering method, which can collect light radiated into the slab and scattered upwards near the waveguide. We therefore use the measurements to locate the loss minimum and the loss floor near the BIC, not to extract radiation losses for \(w < 2\)~\textmu m.

\subsection{Wavelength dependence and design tolerance}\label{sec:wavelength-dependence-and-design-toleran}

Because both effective indices disperse, the BIC follows a continuous line in the \((\lambda,w)\) plane, shifting from \(w \approx 2.1\)~\textmu m at \(\lambda = 1.4\)~\textmu m to 2.3~\textmu m at 1.6~\textmu m [Fig.~\ref{fig:2}(d)]. The BIC width grows more slowly than \(\lambda\) because \(n_{\mathrm{TE}}^{2} - n_{\mathrm{TM}}^{2}\) decreases towards shorter wavelengths as both modes become more strongly confined, partly compensating the wavelength scaling of \(2\pi/k_{x,c}\). The region with radiation loss below 1~dB/cm has an approximately constant width of 0.3~\textmu m, so a single ridge width near 2.2~\textmu m keeps radiation below 1~dB/cm over a bandwidth of about 200~nm and tolerates width errors of about ±0.15~\textmu m. This tolerance, which reflects the weak structural dispersion of a mode confined mainly in the unpatterned film~\cite{yu2019}, is compatible with direct laser writing.

An effective-index analysis (Methods) reproduces these trends. It reproduces the modal hierarchy, with the guided-mode index between the TM slab index of the bare film and \(n_{\mathrm{TE}}\) at every width, and gives \(n_{\mathrm{TM}} = 1.756\) at \(w = 2.25\)~\textmu m, within 0.8\% of the full-vectorial value. However, it underestimates the small difference \(n_{\mathrm{TE}}^{2} - n_{\mathrm{TM}}^{2}\) (0.35 instead of 0.42) on which the BIC width depends. Effective-index results are therefore used only for envelopes, trends and relative sensitivities, and absolute BIC positions are taken from the full-vectorial calculations. Normalized to its value at 1550~nm, the BIC dispersion of Fig.~\ref{fig:2}(d) is reproduced by the effective-index model to within about 1\% [Fig.~\ref{fig:2}(e)]. The sensitivity of the BIC width to the refractive indices and thicknesses of the two layers, obtained with the same model, is discussed in Section 3.3.

\subsection{The BIC in a real device}\label{sec:the-bic-in-a-real-device}

In a real device, every BIC is approached only approximately [Fig.~\ref{fig:2}(f)]. Near \(w_{m}\), the interference factor grows as \(\left( k_{x,c}\delta w/2 \right)^{2}\) with the width detuning \(\delta w = w - w_{m}\), so the radiation-limited quality factor \(Q_{\mathrm{rad}} = 2\pi n_{\mathrm{g}}/\left( \lambda\alpha_{\mathrm{rad}} \right)\), with \(n_{\mathrm{g}}\) the group index, diverges as \(\delta w^{- 2}\). Residual detuning, profile asymmetry or roughness turns the ideal BIC into a quasi-BIC with large but finite \(Q_{\mathrm{rad}}\). In a real device, \(Q_{\mathrm{int}}^{- 1} = Q_{\mathrm{rad}}^{- 1} + Q_{\mathrm{non-rad}}^{- 1}\), where \(Q_{\mathrm{non-rad}}\) accounts for absorption and scattering, so the intrinsic quality factor saturates at the non-radiative limit near the BIC.

Figure~\ref{fig:2}(g) shows the effect of the BIC in the film plane, using a ray model based on Eq.~(\ref{eq:loss}). Away from the BIC (\(w = 3.3\)~\textmu m), the guided mode feeds two wedges of TE slab waves that leave the ridge at \(\theta \approx 20^{\circ}\), and only 3.2\% of the guided power remains after 10~mm. At the BIC width the radiated intensity is suppressed by more than three orders of magnitude, and at least 99.8\% remains.

\subsection{Radius selection and coupler design}\label{sec:radius-selection-and-coupler-design}

In a bend, the TM-like guided mode shifts towards the outer sidewall and the mirror symmetry of the cross-section is lost. The two sidewall contributions then differ in amplitude and acquire an additional relative phase, so the straight-waveguide BIC width no longer ensures cancellation. Rotational symmetry nevertheless preserves the essential structure of the problem. At fixed azimuthal order \(q\), the TE slab continuum consists of a single family of cylindrical waves, regular at the ring center and outgoing at large radius; the emission towards the interior is not a separate channel but part of the same standing-wave pattern. The radiation amplitude can be viewed as the interference of an outer- and an inner-sidewall term with unequal weights~\cite{zou2015,yu2019}, and its cancellation remains a single real condition. Curvature therefore does not destroy the BIC but transforms the discrete BIC widths into BIC bands in the \((R,w)\) plane.

\begin{figure}[!tbp]
  \centering
  \includegraphics[width=\textwidth]{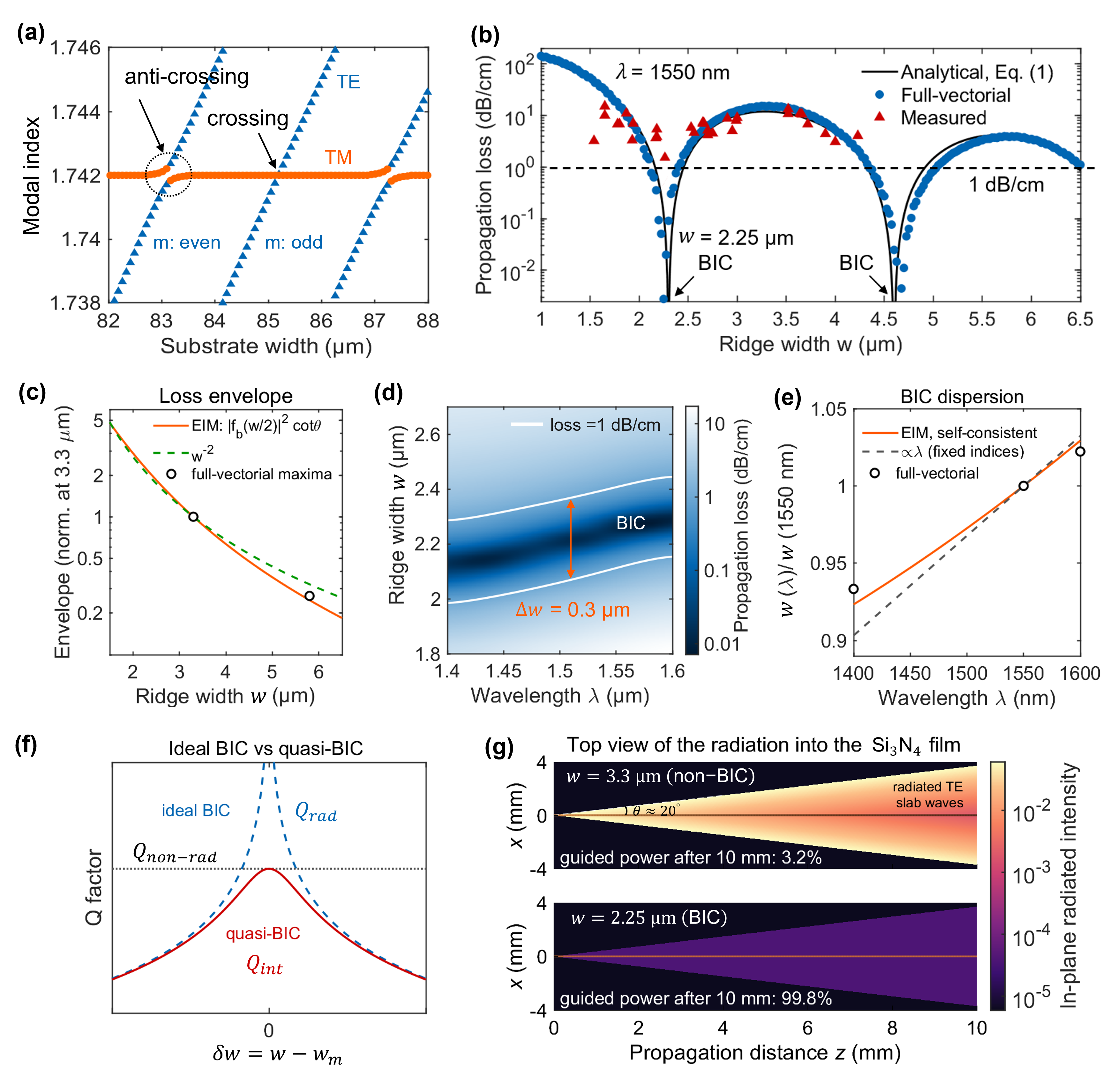}
  \caption{\captiontitle{Width selection of the BIC and its behavior in a real device.} \textbf{(a)} Effective indices of the TM-like guided mode (orange) and of discrete TE modes of a laterally finite slab (blue) versus slab width. The even TE branch anti-crosses the guided mode, showing that the even channel is coupled; the odd TE branch, of opposite symmetry, crosses it and takes no part in the radiation. \textbf{(b)} Propagation loss of the TM-like guided mode versus ridge width at \(\lambda = 1550\)~nm: analytical model [Eq.~(\ref{eq:loss}), line], full-vectorial radiation loss (circles) and measured total propagation loss of laser-written straight waveguides (triangles, top-scattering method). BICs occur at \(w \approx 2.25\) and 4.6~\textmu m; the dashed line marks 1~dB/cm. \textbf{(c)} Envelope \(\left| f_{b}(w/2) \right|^{2}\cot\theta\) of Eq.~(\ref{eq:loss}), where \(f_{b}\) is the lateral profile of the guided mode from the effective-index method (EIM), normalized at 3.3~\textmu m and compared with a \(w^{- 2}\) law and with the full-vectorial loss maxima of (b). \textbf{(d)} Radiation loss versus wavelength and width; white contours enclose the region below 1~dB/cm (\(\Delta w \approx 0.3\)~\textmu m). \textbf{(e)} First BIC width versus wavelength, normalized to 1550~nm: self-consistent EIM solution, pure wavelength scaling at fixed indices, and full-vectorial values from (d). \textbf{(f)} Ideal BIC versus quasi-BIC (schematic): \(Q_{\mathrm{rad}}\) diverges as \(\delta w^{- 2}\) at the interference condition \(w = w_{m}\), whereas the intrinsic quality factor \(Q_{\mathrm{int}}\) of a real device saturates at the non-radiative limit \(Q_{\mathrm{non-rad}}\). \textbf{(g)} Top view of the radiation into the Si\textsubscript{3}N\textsubscript{4} film from a ray model of Eq.~(\ref{eq:loss}), for a leaky width (\(w = 3.3\)~\textmu m) and the BIC width (\(w = 2.25\)~\textmu m), with losses taken from (b). TE slab waves leave the ridge at \(\theta = \arccos\left( n_{\mathrm{TM}}/n_{\mathrm{TE}} \right) \approx 20^{\circ}\); the guided power remaining after 10~mm is indicated. Full-vectorial simulations include radiation only; EIM calculations use non-dispersive materials and a semi-infinite SiO\textsubscript{2} buffer.}
  \label{fig:2}
\end{figure}

\begin{figure}[!tbp]
  \centering
  \includegraphics[width=\textwidth]{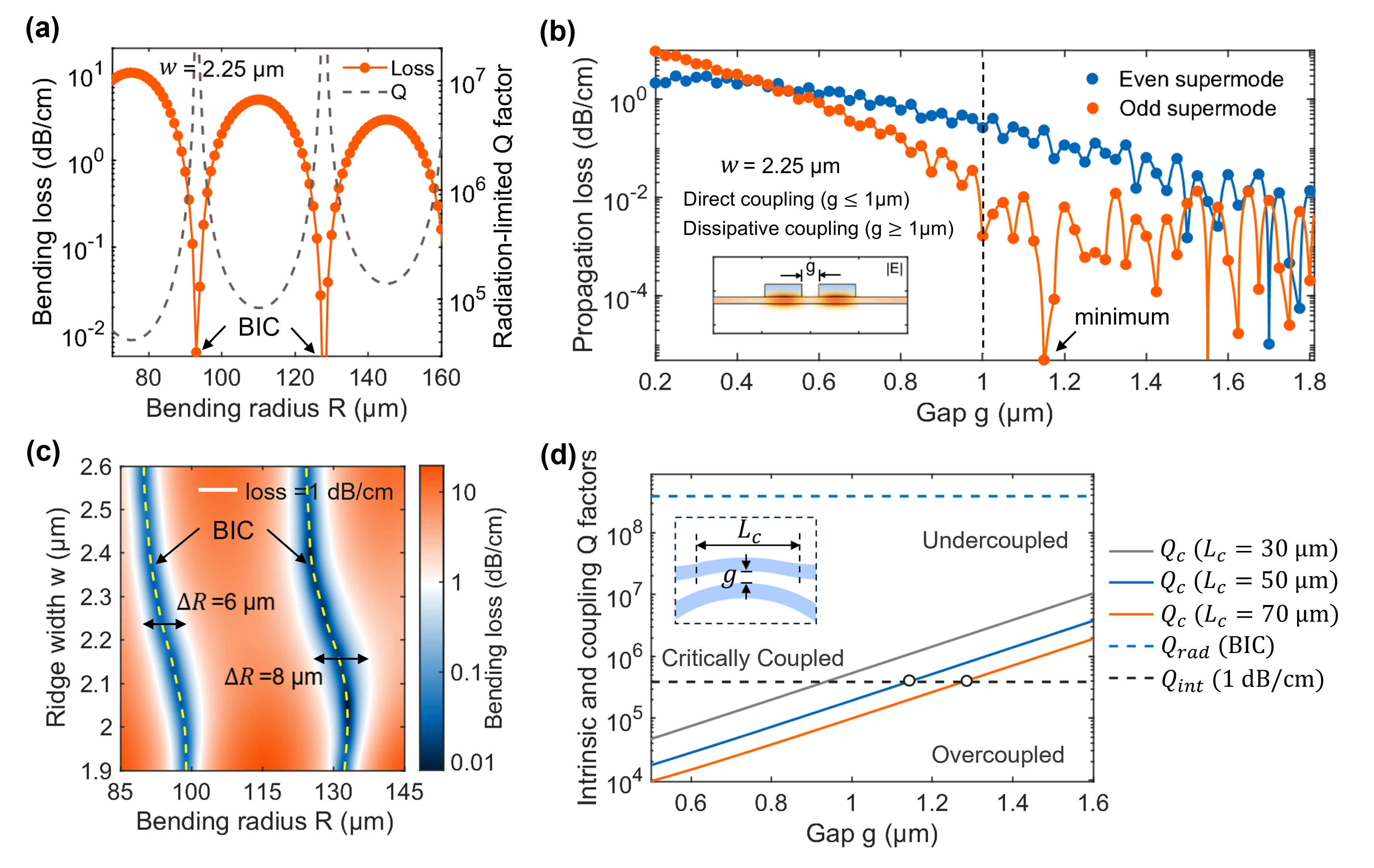}
  \caption{\captiontitle{Radius selection and coupler design.} \textbf{(a)} Simulated bending loss (circles, left axis) and radiation-limited quality factor \(Q_{\mathrm{rad}} = 2\pi n_{\mathrm{g}}/\left( \lambda\alpha_{\mathrm{rad}} \right)\) (dashed, right axis) versus ring radius for \(w = 2.25\)~\textmu m, showing curved BICs at \(R = 93\) and 128~\textmu m. \textbf{(b)} Simulated radiation loss of the even and odd supermodes of two parallel ridges (\(w = 2.25\)~\textmu m) versus edge-to-edge gap \(g\); the dashed line at \(g = 1\)~\textmu m separates the regimes dominated by direct and by dissipative (continuum-mediated) coupling. Inset: \(|E|\) of a supermode. The odd-supermode minimum near \(g \approx 1.15\)~\textmu m lies below the numerical accuracy of the simulation (about \(10^{- 4}\)--\(10^{- 3}\)~dB/cm) and is not resolved. \textbf{(c)} Bending loss versus radius and width. Dashed lines follow the loss minima (curved BICs); white contours mark 1~dB/cm and delimit low-loss bands of width \(\Delta R \approx 6\) and 8~\textmu m. \textbf{(d)} Coupling quality factor \(Q_{\mathrm{c}}\) versus gap for effective coupling lengths \(L_{\mathrm{c}} = 30\), 50 and 70~\textmu m (inset), compared with the radiation limit at the curved BIC, \(Q_{\mathrm{rad}}\), and with \(Q_{\mathrm{int}} \approx 3.9 \times 10^{5}\), corresponding to 1~dB/cm. Circles mark critical coupling; \(n_{\mathrm{g}} = 2.21\).}
  \label{fig:3}
\end{figure}

The simulated bending loss at \(w = 2.25\)~\textmu m illustrates this behavior [Fig.~\ref{fig:3}(a)]. Instead of decreasing monotonically with \(R\), as for an index-guided bend, the loss oscillates: it is about 10~dB/cm at \(R = 80\)~\textmu m, reaches local maxima of about 5 and 3~dB/cm near \(R \approx 111\) and 145~\textmu m, and drops below 0.01~dB/cm at \(R = 93\) and 128~\textmu m. The recurrence of these curved BICs, about 35~\textmu m apart, indicates that the curvature-induced imbalance between the sidewall contributions is periodically compensated as the radius, and with it the azimuthal order, increases. The corresponding radiation-limited quality factor exceeds \(10^{7}\) at the curved BICs and falls to \(9 \times 10^{4}\) and \(1.4 \times 10^{5}\) between them. At fixed cross-section, the choice of radius alone therefore changes the radiation limit by more than two orders of magnitude, and a straight-waveguide BIC design does not by itself guarantee a high-\(Q\) ring.

In the coupling region, the ring and bus form a two-waveguide system with even and odd supermodes. In this platform the coupler is also a radiating structure, because both waveguides couple to the same TE slab continuum, which mediates a dissipative coupling in addition to the direct, evanescent one~\cite{feng2023,gu2023}. Figure~\ref{fig:3}(b) shows the radiation loss of the supermodes of two parallel ridges with \(w = 2.25\)~\textmu m as a function of the edge-to-edge gap. Since each isolated ridge is at its BIC, any supermode radiation originates from the interaction itself. For \(g \lesssim 1\)~\textmu m, direct coupling hybridizes the two ridges and unbalances the sidewall contributions of each, much as curvature does; both supermodes radiate, with losses that decrease from a few dB/cm at \(g = 0.2\)~\textmu m to below 0.3~dB/cm at \(g \approx 1\)~\textmu m. For \(g \gtrsim 1\)~\textmu m, the direct coupling becomes weak and the residual radiation of the two ridges interferes through the shared continuum with a gap-dependent phase. The two supermodes then acquire complementary losses, and the odd supermode develops a pronounced minimum near \(g \approx 1.15\)~\textmu m, consistent with a Friedrich--Wintgen-type supermode BIC~\cite{feng2023,fw1985}. Its depth, like the fine oscillations of both curves below about 10\textsuperscript{\textminus3} dB/cm, lies below the numerical accuracy of the simulation, so it is labeled a minimum and not interpreted quantitatively. The conclusion relevant to the design is robust: for \(g \geq 0.8\)~\textmu m the coupler adds radiation well below the 1~dB/cm floor of the fabricated waveguides. The \((R,w)\) map in Fig.~\ref{fig:3}(c) shows the curved BICs as two continuous low-loss bands, as expected from the single-channel argument above. Their position shifts by only a few micrometers in \(R\) as \(w\) varies from 1.9 to 2.6~\textmu m, and their widths, defined by a bending loss below 1~dB/cm, are \(\Delta R \approx 6\)~\textmu m at 93~\textmu m and \(\Delta R \approx 8\)~\textmu m at 128~\textmu m. We selected \(R = 128\)~\textmu m for fabrication because this band is wider and less sensitive to \(w\).

The saturation of \(Q_{\mathrm{int}}\) at the non-radiative limit [Fig.~\ref{fig:2}(f)] determines the coupler design. The external coupling follows from the supermode beat. With \(\kappa_{s} = \left( \beta_{\mathrm{even}} - \beta_{\mathrm{odd}} \right)/2\) and an effective coupling length \(L_{\mathrm{c}}\), the power coupled out per round trip is \(K = \sin^{2}\left( \kappa_{s}L_{\mathrm{c}} \right)\), which gives the coupling quality factor \(Q_{\mathrm{c}} = 2\pi n_{\mathrm{g}}L/(\lambda K)\) for a ring of circumference \(L = 2\pi R\). The loaded quality factor satisfies \(Q_{\mathrm{load}}^{- 1} = Q_{\mathrm{int}}^{- 1} + Q_{\mathrm{c}}^{- 1}\), and critical coupling requires \(Q_{\mathrm{c}} = Q_{\mathrm{int}}\). Figure~\ref{fig:3}(d) shows \(Q_{\mathrm{c}}\) for \(L_{\mathrm{c}} = 30\), 50 and 70~\textmu m, rising by roughly two decades per micrometer of gap, together with two reference intrinsic values: the radiation limit at the curved BIC, \(Q_{\mathrm{rad}} \approx 4 \times 10^{8}\), and \(Q_{\mathrm{int}} \approx 3.9 \times 10^{5}\), which corresponds to a 1~dB/cm loss floor. For the latter, critical coupling occurs at \(g \approx 0.94\), 1.15 and 1.28~\textmu m for the three coupling lengths. The three-decade separation between these two references is the central design consequence of the BIC: once radiation is cancelled, the coupler must be matched to the non-radiative loss, not to the radiation limit, and designing for \(Q_{\mathrm{rad}}\) would leave the ring strongly under-coupled.

\subsection{Laser-written microring resonators}\label{sec:laser-written-microring-resonators}

The characterized microring [Fig.~\ref{fig:4}(a)] has \(w = 2.3\)~\textmu m, \(R = 128\)~\textmu m, an edge-to-edge gap \(g \approx 0.8\)~\textmu m measured by scanning electron microscopy (SEM) and \(L_{\mathrm{c}} = 70\)~\textmu m. The broadband spectrum [Fig.~\ref{fig:4}(b)] shows a single regular resonance family between 1515 and 1602~nm with a free spectral range (FSR) of about 171~GHz (1.35~nm at 1540~nm). The corresponding group index, \(n_{\mathrm{g}} = c/(2\pi R \cdot FSR) \approx 2.18\), agrees within 1\% with the simulated value of 2.21 for the TM-like guided mode, and no second family is resolved, which identifies the resonances with the designed mode. In both measurements, TM-polarized light was launched through a polarization controller and a polarization-maintaining lensed fiber, in the configurations of Fig.~\ref{fig:4}(c) for the broadband survey and Fig.~\ref{fig:4}(d) for the tunable-laser measurement, which also shows the end-fire set-up. Because the 7.5~GHz resolution of the spectrum analyzer is comparable to the linewidths, this spectrum was used only to locate resonances; linewidths and extinctions were measured with a tunable laser.

Two resonances measured in this way are shown in Fig.~\ref{fig:4}(e), (f). The resonance at 1543~nm has a linewidth of 12~GHz, giving \(Q_{\mathrm{load}} = \nu_{0}/\Delta\nu \approx 1.6 \times 10^{4}\), and an extinction of 0.9~dB; the resonance at 1536~nm has a linewidth of 16.7~GHz (\(Q_{\mathrm{load}} \approx 1.2 \times 10^{4}\)) and a deeper extinction. For an all-pass ring with normalized on-resonance transmission \(T_{0} = \left\lbrack \left( Q_{\mathrm{int}}^{- 1} - Q_{\mathrm{c}}^{- 1} \right)/\left( Q_{\mathrm{int}}^{- 1} + Q_{\mathrm{c}}^{- 1} \right) \right\rbrack^{2}\), the intrinsic quality factor follows as \(Q_{\mathrm{int}} = 2Q_{\mathrm{load}}/\left( 1 \pm \sqrt{T_{0}} \right)\), with the upper (lower) sign for over- (under-) coupling. The transmission spectrum is invariant under exchange of \(Q_{\mathrm{int}}\) and \(Q_{\mathrm{c}}\), so the regime must be assigned independently. Two arguments favor over-coupling. First, the coupler model gives \(Q_{\mathrm{c}} \approx 4 \times 10^{4}\) at \(g = 0.8\)~\textmu m and \(L_{\mathrm{c}} = 70\)~\textmu m, below any intrinsic value compatible with the straight-waveguide loss floor. Second, the two assignments imply very different losses. Over-coupling gives \(Q_{\mathrm{int}} \approx 3.3 \times 10^{5}\) and \(Q_{\mathrm{c}} \approx 1.7 \times 10^{4}\) at 1543~nm, and \(Q_{\mathrm{int}} \approx 1.3 \times 10^{5}\) at 1536~nm, corresponding through \(\alpha = 2\pi n_{\mathrm{g}}/\left( \lambda Q_{\mathrm{int}} \right)\) to 1.2 and 2.9~dB/cm. The value at 1543~nm is comparable to the loss floor measured independently on straight waveguides near the BIC width. Under-coupling would give \(Q_{\mathrm{int}} \approx 1.7 \times 10^{4}\) and \(1.3 \times 10^{4}\), that is 23--30~dB/cm, more than an order of magnitude above that floor.

Figure~\ref{fig:4}(g) summarizes the argument. The measured \(Q_{\mathrm{load}}\) and \(T_{0}\) restrict \(\left( Q_{\mathrm{int}},Q_{\mathrm{c}} \right)\) to two points, related by exchange of the two quality factors. Only the over-coupled point is compatible with both independent constraints: the coupler model at the measured gap and the loss floor of straight waveguides. The extracted \(Q_{\mathrm{c}}\) is about a factor of two below the modeled value, a difference equivalent to a gap error of about 0.15~\textmu m or a longer effective coupling length, both within the uncertainty of the coupler geometry. The over-coupled assignment is therefore the self-consistent one, but an unambiguous determination requires rings with a series of gaps or a phase-sensitive measurement. The latter discriminates directly between the two solutions: although their transmitted intensities are identical, the transmission phase winds by \(2\pi\) across the resonance only in the over-coupled case and shows only a small excursion in the under-coupled case.

\section{Discussion}\label{sec:discussion}

\subsection{Etch-free fabrication and the physical origin of loss}\label{sec:etch-free-fabrication-and-the-physical-o}

In etched Si\textsubscript{3}N\textsubscript{4} waveguides, well-established processes control scattering at the patterned sidewalls through lithography, etch chemistry, sidewall smoothing and annealing, and reach very low losses~\cite{liu2021,pfeiffer2018}. Scattering from a rough boundary scales with the square of the permittivity step across it, the square of the roughness amplitude and the optical intensity at the boundary, and these three factors provide a useful frame for the present platform. Here the nitride film is not patterned at all: it is used as deposited, and lateral confinement is produced by an SU-8 ridge defined in a single maskless direct-laser-lithography exposure, building on our laboratory's previous work on laser-written SU-8 microring lasers and resonators~\cite{sorayaie,hajshahvaladi2025,kavousi2025}. Scattering associated with patterned nitride sidewalls therefore does not arise, while the radiative loss into the film is set by the ridge geometry and engineered through the BIC condition. The process needs no mask, hard mask or dry-etch tool, and a geometry can be changed between writing runs, which is how the width series of Fig.~\ref{fig:2}(b) and the radius selection of Fig.~\ref{fig:3}(c) were realized.

\begin{figure}[!tbp]
  \centering
  \includegraphics[width=\textwidth]{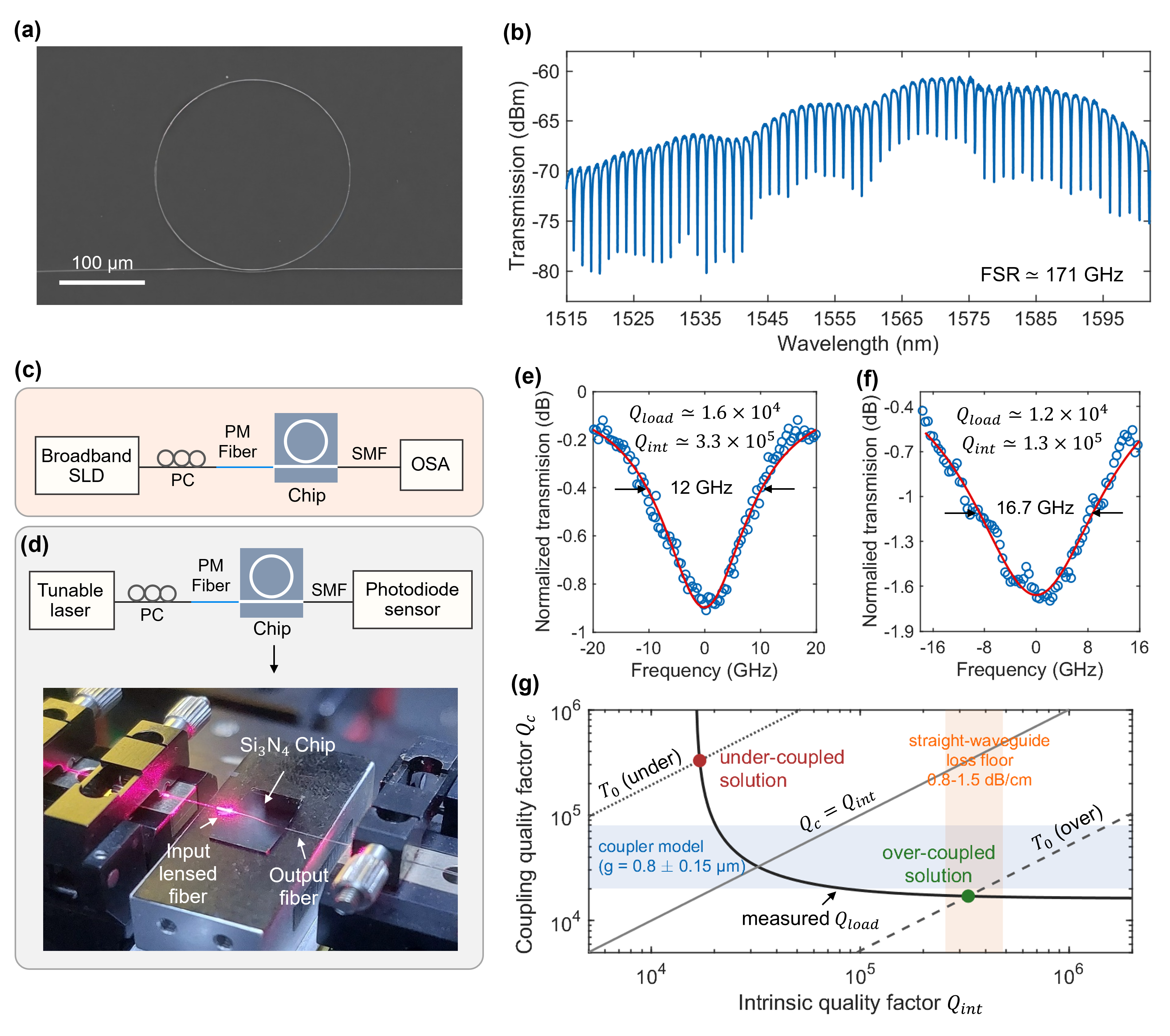}
  \caption{\captiontitle{Laser-written microring resonator on a radius-selected BIC band.} \textbf{(a)} SEM image of the fabricated SU-8 microring side-coupled to a straight SU-8 bus waveguide on the Si\textsubscript{3}N\textsubscript{4} film. \textbf{(b)} Broadband transmitted power of the ring (\(w = 2.3\)~\textmu m, \(R = 128\)~\textmu m, \(g \approx 0.8\)~\textmu m) measured with a superluminescent diode and an optical spectrum analyzer, not normalized to the source spectrum; FSR \(\approx 171\)~GHz, corresponding to \(n_{\mathrm{g}} \approx 2.18\). \textbf{(c)} Configuration for the broadband survey. \textbf{(d)} Configuration for the tunable-laser linewidth measurement, with a photograph of the end-fire coupling set-up with input and output lensed fibers. SLD, superluminescent diode; PC, polarization controller; PM, polarization-maintaining; SMF, single-mode fiber; OSA, optical spectrum analyzer. \textbf{(e)}, \textbf{(f)} Normalized transmission of the resonances at 1543 and 1536~nm measured with the tunable laser (circles) with Lorentzian fits (lines). The linewidths of 12 and 16.7~GHz give \(Q_{\mathrm{load}} \approx 1.6 \times 10^{4}\) and \(1.2 \times 10^{4}\); the intrinsic quality factors (\(3.3 \times 10^{5}\) and \(1.3 \times 10^{5}\)) under the over-coupled assignment. \textbf{(g)} Coupling-regime analysis of the 1543~nm resonance. The measured \(Q_{\mathrm{load}}\) (solid) and on-resonance transmission \(T_{0}\) (dashed and dotted) admit two solutions. Shaded bands: coupler model at \(g = 0.8 \pm 0.15\)~\textmu m with \(L_{\mathrm{c}} = 70\)~\textmu m (blue) and straight-waveguide loss floor of 0.8--1.5~dB/cm converted with \(n_{\mathrm{g}} = 2.18\) (orange). Panels (a)--(f) are measured data or set-up schematics; (g) is a calculation based on the measured values.}
  \label{fig:4}
\end{figure}

A sidewall is not eliminated, however; it is moved to the SU-8 ridge, where the conditions are different. The permittivity step at the SU-8/air boundary (about 1.5) is 1.7--2.4 times smaller than at a Si\textsubscript{3}N\textsubscript{4} boundary to SiO\textsubscript{2} or air (2.5--3.6), which reduces the squared contrast that enters the scattering rate by a factor of about 3--6. In addition, the TM-like guided mode is concentrated in the nitride film beneath the ridge [Fig.~\ref{fig:1}(b), (e)], so the field at the polymer sidewall is weak. These two factors favor the hybrid geometry, but the third does not: laser-written SU-8 sidewalls are shaped by the exposure voxel, stage motion and development, their roughness has not been characterized here, and it is likely larger than that achieved by optimized nanofabrication. Roughness also has a consequence specific to this architecture. Because the ridge sidewalls are the sources of polarization conversion, random roughness perturbs the two-sidewall interference locally and scatters part of the guided mode into the TE slab continuum with random phase, a leakage that the BIC condition of the ideal, translationally or rotationally invariant structure does not cancel.

Scattering from inhomogeneities distributed through the guiding region also contributes. The relevant scatterers are density and composition fluctuations in the deposited nitride film, which carries most of the modal energy; the top and bottom surfaces of the film, whose roughness is set by deposition; the SU-8/Si\textsubscript{3}N\textsubscript{4} interface, where residues or incomplete adhesion can introduce index perturbations; and spatial variations of the cross-link density in the laser-exposed SU-8. Because these scatterers are distributed over the mode rather than concentrated at an edge, their contribution scales with the fraction of modal energy in each material.

Optical absorption is the third non-radiative channel. Deposited nitride films contain residual N--H bonds whose overtone absorption near 1520~nm contributes to loss in the telecom band, with a magnitude that depends on deposition and annealing~\cite{liu2021,pfeiffer2018}. SU-8, like other polymers, absorbs through vibrational overtones of C--H and O--H bonds in the near infrared, weighted by the fraction of modal power in the ridge. Absorption is independent of the BIC condition and therefore sets a floor that geometry cannot remove.

Radiation, finally, is the channel that the design controls. Its role is established experimentally: away from the BIC width the measured losses follow the simulated radiation loss, and the loss minimum coincides with the predicted BIC [Fig.~\ref{fig:2}(b)]. The residual radiation of the fabricated geometry is inferred from simulation: the fabricated width (2.3~\textmu m) lies within the window where the simulated radiation loss is below 1~dB/cm [Fig.~\ref{fig:2}(d)], and \(R = 128\)~\textmu m lies within the corresponding curved-BIC band [Fig.~\ref{fig:3}(c)]. In summary, radiation dominates the loss away from the BIC, whereas near the BIC the total loss is approximately 1 to 1.5~dB/cm in straight waveguides and about 1.2~dB/cm in the ring under the over-coupled assignment. The partition of this non-radiative floor among absorption, volume scattering and sidewall scattering is not resolved by the present measurements; it could be addressed by spectrally resolved loss measurements, annealing of the nitride film and roughness metrology of the laser-written sidewalls.

\subsection{Loss hierarchy and the meaning of the BIC in a fabricated device}\label{sec:loss-hierarchy-and-the-meaning-of-the-bi}

At the curved BIC, the radiation limit is \(Q_{\mathrm{rad}} \approx 4 \times 10^{8}\); for \(Q_{\mathrm{int}} = 3.3 \times 10^{5}\), radiation therefore accounts for less than 0.1\% of the intrinsic loss rate. Without the BIC, the same ring would be limited to \(Q_{\mathrm{rad}} \approx 10^{5}\) at radii between the curved BICs [Fig.~\ref{fig:3}(a)], and straight waveguides away from the BIC width would lose several to tens of dB/cm [Fig.~\ref{fig:2}(b)]. The design thus removes radiation as a limiting channel, and further improvement of \(Q_{\mathrm{int}}\) becomes a materials and fabrication problem rather than an electromagnetic one. This hierarchy also fixes the terminology. In the model, the TM-like guided mode at the BIC width or radius is an exact BIC with respect to the TE slab continuum. In a fabricated device, any deviation from that condition, such as the difference between the fabricated (2.3~\textmu m) and simulated (2.25~\textmu m) widths, a non-rectangular ridge profile or sidewall roughness, restores a small radiative coupling and turns the mode into a quasi-BIC with large but finite \(Q_{\mathrm{rad}}\) [Fig.~\ref{fig:2}(f)]~\cite{hsu2016,dong2022}. Absorption and volume scattering are separate, non-radiative channels that exist whether or not the BIC condition is met. The measured \(Q\) values are limited by these channels and therefore provide a lower bound on \(Q_{\mathrm{rad}}\), not a measurement of it.

\subsection{Design strategy, geometry control and practical constraints}\label{sec:design-strategy-geometry-control-and-pra}

Each structural parameter controls a distinct physical quantity. The film and ridge thicknesses set the effective indices and hence the lateral wavenumber of the continuum; the ridge width sets the interference phase between the two sidewall sources; the bending radius sets the amplitude imbalance and the additional phase introduced by curvature; and the gap and coupling length set the external coupling and the radiation of the supermodes. The design therefore proceeds hierarchically: the width is fixed at a straight-waveguide BIC, the radius is chosen on a curved-BIC band at that width, and the coupler is matched to the realistic non-radiative loss rather than to the radiation limit [Fig.~\ref{fig:3}(d)]. This hierarchy distinguishes a BIC resonator from a ring that merely uses a straight-BIC width: at \(w = 2.25\)~\textmu m, a radius between the curved BICs would limit \(Q_{\mathrm{rad}}\) to about \(10^{5}\) [Fig.~\ref{fig:3}(a)], below the intrinsic \(Q\) measured here.

The parameters do not act independently. The curved-BIC radius depends on the width [Fig.~\ref{fig:3}(c)], the BIC width depends on wavelength [Fig.~\ref{fig:2}(d)] and on the film and ridge properties, and the supermode radiation depends on both width and gap [Fig.~\ref{fig:3}(b)]. Moreover, the loss is an oscillatory function of each parameter [Figs.~\ref{fig:2}(b), 3(a) and 3(b)], so the loss landscape consists of narrow minima separated by maxima one to three orders of magnitude higher, and sequential one-parameter optimization can converge on minima that are not robust. We therefore mapped the loss over the \((\lambda,w)\) and \((R,w)\) planes and selected the operating point by its tolerance as well as its depth.

These sensitivities make geometry control by direct laser lithography the main lever on device performance. The written linewidth is set by the exposure dose, through the laser power and writing speed, and by focus and resist thickness. Because the ridge width fixes the interference phase, a width error translates directly into residual radiation; across the measured width series the loss changes by more than an order of magnitude over a few hundred nanometers [Fig.~\ref{fig:2}(b)]. The low-loss windows, about ±0.15~\textmu m in width and a few micrometers in radius, are compatible with the present process, as the fabricated ring (\(w = 2.3\)~\textmu m, \(R = 128\)~\textmu m) lies within both. The writing process also determines sidewall quality, cross-link uniformity and the start and end point of the written ring. This closure point is a local defect that breaks rotational symmetry and may locally degrade the cancellation; the stage-acceleration control described in Methods was introduced to mitigate it. Improved dose calibration, focus control and trajectory control at the ring closure are expected to reduce geometrical imperfections and the associated losses, but such improvements have not yet been quantified, and reproducibility across devices and larger areas requires statistical characterization beyond the single ring analyzed here. Because the BIC mode is TM-like and both the continuum and the TE-like ridge mode have the orthogonal polarization, the input polarization must also be well controlled: residual TE excitation adds a background that distorts the measured extinction.

Finally, the refractive indices of SU-8 and Si\textsubscript{3}N\textsubscript{4} were not measured on the present samples. Because \(n_{\mathrm{TE}}^{2} - n_{\mathrm{TM}}^{2} \approx 0.42\) is a small difference of two large quantities, the BIC width is sensitive to any differential shift of the two effective indices. According to the effective-index analysis (Methods), a 1\% change of the Si\textsubscript{3}N\textsubscript{4} index shifts the first BIC by about 4--5\% (about 0.1~\textmu m, comparable to the ±0.15~\textmu m tolerance window), and a 1\% change of the SU-8 index by about 3\%. Film and ridge thicknesses change \(n_{\mathrm{TE}}\) and \(n_{\mathrm{TM}}\) almost in common mode: a 5\% change of either thickness shifts the BIC width by less than 1\%, and the ridge height enters only weakly because the guided field is evanescent in the SU-8 (\(n_{\mathrm{TM}} > n_{\mathrm{SU8}}\)). The agreement of the measured loss minimum with the predicted BIC width, and of the measured group index with simulation, indicates that the assumed indices are adequate, but a quantitative uncertainty analysis will require ellipsometry of the films.

\subsection{Applicability, limitations and outlook}\label{sec:applicability-limitations-and-outlook}

The platform trades performance for simplicity, and its range of application should be stated plainly. The measured loss floor of about 1--1.5~dB/cm and the intrinsic quality factor of \(3.3 \times 10^{5}\) remain well below the performance of mature Si\textsubscript{3}N\textsubscript{4} processes~\cite{pfeiffer2018,pfeiffer2016,puckett}, so the approach is not intended for applications that require the lowest loss. Guidance is restricted to the TM-like mode, which requires careful polarization control at the input. The weak lateral confinement provided by the polymer ridge requires bend radii of the order of 100~\textmu m and makes the mode sensitive to nearby structures, which limits integration density. The polymer ridge also restricts subsequent high-temperature processing. Within these limits, the approach offers a short path from design to device, with spin coating and a single maskless exposure, and is suited to prototyping resonator and coupler layouts, to laboratory-scale studies of BIC physics in resonators, and to devices whose loss budget is set by other elements.

Beyond waveguides and rings, the principle is general. Any nanostructure that can be defined as a pattern in a low-index layer on an unetched functional film, such as coupled-resonator arrays, gratings or periodic lattices, couples to the same slab continuum, and its radiation can be engineered by the same interference design. The approach is also not specific to Si\textsubscript{3}N\textsubscript{4}: hybrid BIC waveguides have been realized on lithium niobate, two-dimensional materials and rare-earth oxides~\cite{yu2019,yu2019adom,yu2021adom,xu2021,li2022}. Combined with maskless laser lithography, BIC-guided design could therefore provide a way to prototype such nanostructures on films that are difficult or undesirable to pattern. These extensions have not been demonstrated here, and each geometry requires its own full-vectorial radiation analysis.

The intrinsic \(Q\) values at 1543 and 1536~nm differ by a factor of about 2.5 over 7~nm. This difference may reflect spectral detuning from the curved BIC, the short-wavelength N--H absorption band, wavelength-dependent coupling or uncertainties in the fit and normalization, and the present data do not distinguish these. Because the curved BIC is defined at a single wavelength for a fixed \(R\), the intrinsic \(Q\) of successive resonances should peak where the ring crosses the BIC band. Extracting \(Q_{\mathrm{int}}(\lambda)\) over the full tuning range would test this prediction and provide a direct spectral signature of the BIC in the resonator, analogous to the width-resolved minimum observed in straight waveguides.

A further design direction concerns the ridge material. Replacing SU-8 by SiO\textsubscript{2} would make the index above the nitride film beneath the ridge equal to that of the buffer, so the vertical index profile becomes more symmetric. In a vertically symmetric slab, TE and TM modes have opposite parity, so polarization conversion would arise only from the asymmetry introduced by the ridge itself. This could offer an additional handle on the radiation strength while avoiding the vibrational absorption of the polymer. These benefits are prospective and have not been verified here by simulation or experiment, and the lower index of SiO\textsubscript{2} would further weaken lateral confinement. An oxide ridge would also require deposition and patterning of the SiO\textsubscript{2}, although the nitride itself would remain unpatterned, and would therefore forgo the single-step lithography that is the principal practical advantage of the present process.

\section{Conclusion}\label{sec:conclusion}

We have demonstrated hybrid SU-8/Si\textsubscript{3}N\textsubscript{4} waveguides and microring resonators defined by direct UV laser lithography on an unpatterned nitride film. The approach extends our laboratory's experience with laser-written SU-8 photonic devices to Si\textsubscript{3}N\textsubscript{4}: a single maskless exposure defines the complete waveguide geometry, without patterning the nitride and without additional thin-film deposition. Because the film is not patterned, scattering at patterned nitride sidewalls does not arise, and the central requirement of the platform becomes the control of radiation into the slab continuum, which BIC-guided design provides. Radiation from the two ridge sidewalls cancels by destructive interference at discrete widths. Because a single radiation channel remains, this cancellation is a structurally stable, one-parameter condition that persists across wavelength and, in rings, along bands of radius and width despite curvature. These design rules carried over to fabricated devices. The measured loss of laser-written waveguides follows the predicted width dependence and is lowest, at about 1.5~dB/cm, at the design width. A microring designed on a radius-selected band has a group index consistent with simulation and, under the over-coupled assignment, an intrinsic quality factor of \(3.3 \times 10^{5}\), far below its radiation limit.

These results represent an initial demonstration, and the present losses remain above those of mature Si\textsubscript{3}N\textsubscript{4} processes. The remaining loss is non-radiative, so further improvement is expected from optimizing the writing conditions, dimensional control and material properties rather than from changes to the optical design. More broadly, BIC-guided design offers an etch-free route with fewer fabrication steps, complementary to established nanofabrication, which could be extended to other nanostructures defined on unpatterned films, including coupled-resonator architectures such as photonic molecules and Vernier filters~\cite{hajshahvaladi2025,zhang2021,zhang2019,ji2024}.

\section{Methods}\label{sec:methods}

\subsection{Sample fabrication}\label{sec:sample-fabrication}

The substrate was a silicon wafer covered with a 2~\textmu m SiO\textsubscript{2} layer and a 400~nm LPCVD Si\textsubscript{3}N\textsubscript{4} film. It was cleaned ultrasonically in acetone and isopropyl alcohol (5~min each), rinsed in deionized water, dried with nitrogen and dehydrated on a hot plate at 95~°C and then 110~°C (10~min each). After cooling, SU-8 2001 was spin-coated at 3000~rpm to a thickness of about 800~nm and soft-baked at 95~°C for 2~min. No etching step was used at any stage: the Si\textsubscript{3}N\textsubscript{4} film was used as deposited, and waveguides and microrings were defined in the SU-8 by direct laser lithography~\cite{sorayaie} with a 405~nm laser (20~mW) focused through a 40× objective while the sample was translated on a motorized \(XY\) stage; a mechanical shutter with millisecond switching controlled the exposure. The written linewidth was set by the laser power and writing speed. A writing speed of 40--60~\textmu m/s and a stage acceleration of about 8~mm/s\textsuperscript{2} were used for microrings to improve the positioning of the trajectory at the start and end of each ring. The device geometry was inspected by SEM.

\subsection{Propagation-loss measurement}\label{sec:propagation-loss-measurement}

Propagation losses of straight waveguides were measured by the top-scattered-light method. Light from a tunable laser (Thorlabs TLX1) at 1550~nm was butt-coupled into the waveguide, with initial alignment at 635~nm under a top-view camera. Light scattered from the waveguide was collected by a 100× long-working-distance objective (NA 0.80), spatially filtered by an iris in an intermediate image plane to reject scattering from the input facet and substrate, and detected with a germanium power sensor (Thorlabs S122C). The sample and fiber were translated along the waveguide over 500--1000~\textmu m in steps of 1--2~\textmu m, and the loss was obtained from a linear fit of \(10\log_{10}P_{\mathrm{scat}}(z)\), excluding localized scattering spikes. For losses near 1~dB/cm the total decay over the scan is only 0.05--0.1~dB, so these values carry a larger relative uncertainty than those in the radiation-dominated range.

\subsection{Resonator spectroscopy}\label{sec:resonator-spectroscopy}

Transmission spectra were measured in an end-fire configuration with lensed fibers [Fig.~\ref{fig:4}(c), (d)]. A fiber polarization controller and a polarization-maintaining input fiber prepared TM-polarized input, and the output was collected with a single-mode fiber. Broadband spectra were recorded with a superluminescent diode (Thorlabs S5FC1005S) and a Fourier-transform optical spectrum analyzer (Thorlabs OSA202C; resolution 7.5~GHz). Linewidths were measured by stepping the frequency of the tunable laser (Thorlabs TLX1; 1527.6--1565.5~nm) across each resonance and recording the transmitted power with the germanium sensor. Spectra were normalized to the off-resonance transmission and fitted with a Lorentzian, from which \(Q_{\mathrm{load}} = \nu_{0}/\Delta\nu\) and \(T_{0}\) were obtained.

\subsection{Numerical simulations}\label{sec:numerical-simulations}

Modal properties were computed with a full-vectorial eigenmode solver applied to the waveguide cross-section, using the geometry and indices given above; Fig.~\ref{fig:1}(f) shows \(\mathrm{Re}\left( E_{z} \right)\) of an even TE slab continuum mode of the same cross-section computed with this solver. The domain was terminated by perfectly matched layers, and the radiation loss was obtained from the imaginary part \(n''\) of the complex effective index as \(\alpha_{\mathrm{rad}} = 2k_{0}n''\) (converted to dB/cm). Absorption and scattering were not included, so all simulated losses are radiation losses. The effective-index analysis of Fig.~\ref{fig:2}(c), (e) and of the sensitivity estimates in Section 3.3 used slab indices of the film and ridge regions from a transfer-matrix solver, with non-dispersive materials and a semi-infinite SiO\textsubscript{2} buffer; the lateral TM-like guided mode was then solved as a symmetric three-layer problem with boundary conditions for an electric field parallel to the sidewalls, and the BIC width from the self-consistent condition \(k_{x,c}(w)\, w = 2\pi\). In Fig.~\ref{fig:1}(c), the guided-mode profile at \(w = 2.25\)~\textmu m is taken from the simulation of Fig.~\ref{fig:1}(e); for the narrower ridges, which were not simulated, its width follows the effective-index trend. Figure~\ref{fig:2}(g) uses a ray model: the guided power decays as \(e^{- \alpha_{\mathrm{rad}}z}\), and the power radiated per unit length is carried away by TE slab waves at \(\theta = \arccos\left( n_{\mathrm{TM}}/n_{\mathrm{TE}} \right)\), with propagation loss in the film neglected. Figure~\ref{fig:4}(g) follows from the all-pass coupled-mode relations given in the text, evaluated with the measured \(Q_{\mathrm{load}}\) and extinction.

\bibliographystyle{unsrt-ipl}
\bibliography{references}

\end{document}